\documentclass[a4paper,11pt]{article}

\usepackage{times}
\usepackage{bm}
\usepackage{natbib}

\usepackage{pst-tree}
\usepackage{pstricks}

\usepackage{amssymb,amsmath,amsthm,mathrsfs}
\usepackage{graphicx}

\theoremstyle{plain}  
\newtheorem{theorem}{Theorem}
\newtheorem{lemma}{Lemma}
\newtheorem{corollary}{Corollary}
\newtheorem{definition}{Definition}

\newtheorem{remark}{Remark}

\providecommand{\cov}{\mathrm{cov}}
\providecommand{\var}{\mathrm{var}}
\providecommand{\wcov}{\mathrm{wcov}}

\providecommand{\E}{\mathrm{E}}

\begin{document}




\title{A cautionary note on robust covariance plug-in methods}

\author{Klaus Nordhausen\thanks{Department of Mathematics and Statistics, University of Turku. Email: klaus.nordhausen@utu.fi} \and David E. Tyler\thanks{Department of Statistics, Rutgers University, U.S.A. Email: dtyler@rci.rutgers.edu}}

\maketitle

\begin{abstract}
Many multivariate statistical methods rely heavily on the sample covariance matrix. It is well known though that the sample
covariance matrix is highly non-robust. One popular alternative approach for ``robustifying'' the multivariate method is to simply
replace the role of the covariance matrix with some robust scatter matrix. The aim of this paper is to point out that in
some situations certain properties of the covariance matrix are needed for the corresponding robust ``plug-in'' method to be a valid approach, and that
not all scatter matrices necessarily possess these important properties.  In particular, the following three multivariate methods are discussed
in this paper: independent components analysis, observational regression and graphical modeling. For each case, it is shown that using
a symmetrized robust scatter matrix in place of the covariance matrix results in a proper robust multivariate method.\\

\noindent \textbf{Keywords}: Factor analysis; Graphical model; Independent components analysis; Observational regression, Scatter matrix, Symmetrization.
\end{abstract}


\section{Introduction} \label{Section-Intro}
For a $p$-variate random vector $x = (x_1,\ldots, x_p)^T$ the covariance matrix, or variance-covariance matrix,
\[
\cov(x) =   E \left((x - E(x))(x - E(x))^T \right) = E(xx^T) - E(x)E(x)^T
\]
is a fundamental descriptive measure and is one of the cornerstones in the development of multivariate methods.
The covariance matrix has a number of important basic properties, for example:

\begin{lemma} \label{CovProp}
Let $x$ and $y$ be $p$-variate continuous random vectors with finite second moments, then \\[-20pt]
\begin{enumerate}
  \item The covariance matrix $\cov(x)$ is symmetric and positive semi-definite.
  \item The covariance matrix is affine equivariant in the sense that
  \[
  \cov(Ax+b)= A\cov(x)A^T,
  \]
  for all full rank $p \times p$ matrices $A$ and all $p$-vectors $b$.

  \item If the $i$th and $j$th components of $x$ are independent, then
	\[\left(\cov(x)\right)_{jk}=\left(\cov(x)\right)_{kj}=0.\]
	  \item If x and y are independent, then the covariance matrix is additive in the sense that
  \[
  \cov(x+y)=\cov(x) + \cov(y).
  \]
  \end{enumerate}
\end{lemma}
Furthermore, for a random sample $X_n = (x_1, \ldots, x_n)^T$ coming from a $p$-variate normal distribution $N_p(\mu,\Sigma)$, the finite sample version of $\cov(x)$,
i.e.\ the sample covariance matrix
 \[ S(X_n) = \frac{1}{n} \sum_{i=1}^n (x_i - \bar{x})(x_i - \bar{x})^T \]
is the maximum likelihood estimator for the scatter parameter $\Sigma = \cov(x)$. Also, together with the sample mean vector $\bar{x}$, the sample covariance matrix gives
a sufficient summary of the data under the assumption of multivariate normality. Hence any method derived assuming multivariate normality will be based solely on the the
sample mean vector and sample covariance matrix.

It is well known though that multivariate methods based on the sample mean and sample covariance matrix are highly non-robust to departures from multivariate normality. Such
methods are extremely sensitive to just a single outlier and are highly inefficient at longer tailed distributions. Consequently, a substantial amount of research has been
undertaken in an effort to develop robust multivariate methods which are not based on the mean vector and covariance matrix. A common approach for ``robustifying'' classical multivariate methods based on the sample mean vector and covariance matrix is the ``plug-in'' method, which means to simply modify the method by replacing the mean vector and covariance matrix with robust estimates of multivariate location and scatter. However, sometimes crucial properties of the covariance matrix are needed in order for a
particular multivariate method to be valid, and investigating whether these properties hold for the robust scatter replacement is often not addressed. Typically, scatter
matrices are defined so that they satisfy the first two properties in Lemma~\ref{CovProp}, but not necessarily the other properties.

In this paper, we focus on the third property above and its central role in certain multivariate procedures, in particular in independent components analysis
(section \ref{Section-ICA}), in observational regression (section \ref{Section-ObsReg}) and in graphical modeling (section \ref{Section-graphical}). These cases illustrate why
the use of plug-in methods should be done with some caution since not all scatter matrices necessarily satisfy this property. Some counterexamples are given in
section \ref{Section-Indep}, where it is it also noted that using symmetrized versions of common robust scatter matrices can make the corresponding plug-in method more
meaningful.  Some comments on the computational aspects of symmetrization are made in section \ref{Section-Comp}. All computations
reported in this paper were done using R 2.15.0 \citep{R2150}, and relied heavily on the R-packages ICS \citep{NordhausenOjaTyler:2008}, ICSNP \citep{ICSNP} MASS
\citep{VenablesRipley:2002} and SpatialNP \citep{SpatialNP}. Proofs are reserved for the appendix. To begin, the next section briefly reviews that concepts of scatter
matrices, affine equivariance and elliptical distributions, and sets up the notation used in the paper.

\section{Scatter matrices and affine equivariance} \label{Section-scatter}

Many robust variants of the covariance matrix have been proposed within the statistics literature, with the vast majority of these variants satisfying the following
definition of a scatter, or pseudo-covariance, matrix.
\begin{definition} \label{ScatterDef}
Let $x$ be a $p$-variate random vector with cdf $F_{x}$. A $p \times p$ matrix valued functional $V(F_{x}) = V(x)$ is called a scatter functional
if it is symmetric, positive semi-definite  and affine equivariant in the sense that
\[
  V(A x + b) =   A V(x) A^T,
\]
for any $p \times p$ full rank matrix $A$ and any $p$-vector $b$.
\end{definition}
A scatter statistic $\hat{V}$ is then one that satisfies the above definition when $F_{x}$ is replaced by the empirical cdf.
Scatter statistics which satisfy this definition include M-estimators \citep{Huber:1981,Maronna:1976}, minimum volume ellipsoids (MVE) and minimum covariance determinant (MCD)
estimators \citep{Rousseeuw:1986}, S-estimators \citep{Davies:1987,Lopuhaa:1989}, $\tau$-estimators \citep{Lopuhaa:1991}, projection based scatter estimators
\citep{DonohoGasko:1992,MaronnaStahelYohai:1992,Tyler:1994}, re-weighted estimators \citep{Ruiz-Gazen:1993,Lopuhaa:1999} and MM-estimates \citep{TatsuokaTyler:2000,Tyler:2002}.

Definition~\ref{ScatterDef} emphasizes only the first two properties of the covariance matrix noted in Lemma~\ref{CovProp}, with the other stated properties not
necessarily holding for a scatter functional in general. In addition, a scatter statistic cannot be viewed as an estimate of the population covariance matrix, but rather
as an estimate of the corresponding scatter functional.  For some important distributions, though, a scatter functional and the covariance matrix have a simple
relationship.  For example, elliptically symmetric distributions are often used to evaluate how well a multivariate statistical method performs outside of the normal family.
For such distributions, it is known that if $x$ possesses second moments then  $V(F_x) \propto \cov(x)$.  This relationship also holds for a broader class of
distributions discussed below. We first recall the definition of elliptical distributions \citep[see e.g.][]{BilodeauBrenner:1999}.

\begin{definition}\label{EllDef}
A $p$-variate random vector $y$ is said to be spherically distributed around the origin if and only if $Oy \sim y$ for all orthogonal $p \times p$ matrices $O$.
The random vector $  x $ is said to have an elliptical distribution if and only if it admits the representation $x \sim \Omega y + \mu$ with $y$ having a spherical distribution,
$\Omega$ being a full rank $p \times p$ matrix and $ \mu$ being a $p$-vector.
\end{definition}

If the density of an elliptical distribution exists, then it can be expressed as
\[
f(  x,   \mu,   \Sigma)= |\Sigma|^{-\frac{1}{2}} \exp\left\{ - \rho(||\Gamma^{-1/2}(x - \mu)||_2^2) \right\},
\]
where $\rho(\cdot)$ is a function independent of $\mu$ and $\Gamma$ and $\Gamma = \Omega \Omega^T$. We then say that $x \sim E(\rho, \mu, \Gamma)$.
(For a symmetric positive definite matrix $S$, the notation $S^{1/2}$ refers to its unique symmetric positive semi-definite square root.)
A generalization of the spherical distributions and of the elliptical distributions can be constructed as follows \citep[see][]{Oja:2010}.

\begin{definition}\label{ESSDef}
A $p$-variate random vector $y$ is said to have an exchangeable sign-symmetric distribution about the origin if and only if $P J y \sim y$ for all $p \times p$
permutation matrices $P$ and all $p \times p$ sign-change matrices $J$ (a diagonal matrix with $\pm 1$ on its diagonal).
\end{definition}
The density $f$ (if it exists) of an exchangeable sign-symmetric $y$  must satisfy the property that $f(y)=f(P J y)$ for any $P$ and $J$. We then
denote $x \sim ESS(f, \mu, \Omega)$  if and only if it admits the representation $x \sim \Omega y + \mu$ where $y$ has a exchangeable sign-symmetric distribution
with density $f$,  $\Omega$ is a full rank $p \times p$ matrix and $\mu$ is a $p$-vector. Note that in this model $\Omega$ is not completely identifiable since
$ESS(f, \mu, \Omega) \sim ESS(f, \mu, \Omega^*=\Omega P J)$ for any $P$ and $J$. However, $\Gamma =  \Omega \Omega^T$ is identifiable since
$\Omega^*{\Omega^*}^T=\Omega P J J P^T \Omega^T=\Omega \Omega^T = \Gamma$. On the other hand, unlike the elliptical distributions, the distribution
$ESS(f, \mu, \Omega)$ can not be completely determined from $f, \mu$ and $\Gamma$.

Clearly the multivariate normal distributions are special cases of the family of elliptical distributions and the elliptical distributions in turn belong to the family
of $ESS$ distributions. In particular, $E(\rho, \mu, \Gamma) \sim  ESS(f, \mu, \Gamma^{1/2})$ with $f(y) =  \exp\{ - \rho(y^Ty) \}$. The $ESS$
distributions also contain other well studied distributions such as the family of $L_p$-norm distributions \citep[see for example][]{GuptaSong:1997}.
For $x \sim ESS(f, \mu, \Omega)$ in general, or $x \sim E(\rho, \mu, \Gamma)$ in particular, the parameter $\Gamma \propto \cov(x)$ provided $\cov(x)$ exist, with
the constant of proportionality being dependent on the function $f$ or the function $\rho$ respectively.
To simplify notation, it is hereafter assumed that these functions are standardize so that $\Gamma = \cov(x)$  whenever $x$ which has finite second moments.
If the second moments do not exist, then $\Gamma$ still contains information regarding the linear relationship between the components of $x$.
The following lemma notes that the relationship between $\Gamma$ and $\cov(x)$ extends to any scatter functional.
\begin{lemma} \label{diagVess}
\mbox{  } \\[-20pt]
\begin{enumerate}
\item \indent For any $p$-vector y which is exchangeable sign-symmetric around the origin all scatters matrices are proportional to the identity matrix, i.e.
for any scatter functional $V(y)$ which is well defined at $y$, \\[-20pt]
\[
V(y) = c_f I_p,
\]
where $c_f$ is a constant depending on the density $f$ of $y$.
\item For $x \sim ESS(f, \mu, \Omega)$ with $\Gamma = \Omega\Omega^T$, if the scatter functional $V(x)$ is well-defined at $x$, then \\[-12pt]
\[
V(y) = c_f \Gamma,
\]
where $c_f$ is a constant depending on the function $f$.
\end{enumerate}
\end{lemma}
For these models, all scatter functionals are proportional and so any consistent scatter statistic is consistent for $\Gamma$ up to a
scalar multiple. Consequently, and especially when the function $f$ is not specified for the $ESS(f, \mu, \Omega)$ distribution, the parameter
$\Gamma$ is usually only of interest up to proportionality. This motivates considering the broader class of shape functionals as
defined below. Lemma \ref{diagVess} also holds when $V$ is taken to be a shape functional.
\begin{definition} \label{ShapeDef}
Let $x$ be a $p$-variate random vector with cdf $F_{x}$. Then any $p \times p$ matrix valued functional $V(F_{x}) = V(x)$ is a shape functional
if it is symmetric, positive semi-definite  and affine equivariant in the sense that
\[
  V(A x + b) \propto   A V(x) A^T,
\]
for any $p \times p$ full rank matrix $A$ and any $p$-vector $b$.
\end{definition}
An example of a shape functional which is not a scatter functional is the distribution-free M-estimate of scatter \citep{Tyler:1987}.

It is worth noting that \citet{TylerCritchletDuembgenOja:2009} conjecture in their Remark 1 that the $ESS$ distributions are perhaps the largest class of distributions
which all scatter or shape matrices are proportional to each other. Outside of this class, different scatter or shape statistics estimate different population
quantities. This is not necessarily a bad feature, since as noted by several authors \citep{TylerCritchletDuembgenOja:2009,NordhausenOjaOllila:2011} the comparison of different
scatter/shape matrices can be useful in model selection, outlier detection and clustering.

Note that due to Lemma \ref{diagVess}, any scatter functional satisfies Lemma \ref{CovProp} under an $ESS$ distribution (although properties 3, 4 and 5 are vacuous
for any non-normal elliptical distribution since such distributions do not have any independent components). For general distributions, however, one must check
that the scatter functional used in a plug-in method has the properties of the regular covariance matrix needed for the method at hand.

\section{Independence} \label{Section-Indep}

Although a zero covariance between two variable does not imply the variables are independent, the property that independence implies a zero covariance (when the second
moments exist) is of fundamental importance when one wishes to view the covariance or correlation as a measure of dependency between variables. It has been
pointed out by \citet{OjaSirkiaEriksson:2006} that many of the popular robust scatter matrices do not posses the property, but they do not present any concrete
counterexample. This somewhat surprising observation is not well known and so in this section we explore it in more detail. Some simple counterexamples are given
which not only verify this observation but also demonstrates how large a \emph{pseudo-correlation},
\[ \rho_{jk}\left(V(x)\right) = \frac{V_{jk}(x)}{\sqrt{V_{jj}(x)V_{kk}(x)}},\]
can be even when the corresponding variables are independent.

\subsection{Counterexamples} \label{Counter}
The first example involves the family of weighted covariance matrices, which for a given $\alpha$ is defined as
\[
  \wcov_\alpha(x) =   E \left(r^\alpha (x -   E(x)) (x -   E(x)^T) \right)
\]
where $r= \sqrt{(x -  E(x))^T \cov(x)^{-1} (x -  E(x))}$ is the Mahalanobis distance. It is easy to see that
that $ \wcov_\alpha(x)$ satisfies definition \ref{ScatterDef} for a scatter matrix for $x$ and that it corresponds to the covariance matrix when $\alpha=0$.
The weighted covariance matrices do not necessarily have good robustness properties, especially when $\alpha > 0$ since this corresponds ``up-weighing'' the values of $x$
based on their Mahalanobis distances. They serve, though, as a tractable family of scatter matrices which helps us to illustrate our main points. For simplicity,
assume without loss of generality that $E(x) =  0$ and $\cov(x)=  I_p$, then
\[
  \wcov_\alpha(x) =   E \left((x_1^2 + \ldots + x_p^2)^{\alpha/2} xx^T) \right).
\]
Suppose now that the components of $x$ are mutually independent and consider the case $\alpha=4$. This yields for the diagonal elements
\[
 \{\wcov_4(x)\}_{jj}= E(x_j^6) + 2(p-1)E(x_j^4)+ \sum_{k\neq j} E(x_k^4) + p^2-3p+2
\]
and for the off-diagonal elements
\[
  \{\wcov_4(x)\}_{jk}=2 E(x_j^3) E(x_k^3), 1\leq j \neq k \leq p.
\]
Since $E(x_j^3)$ corresponds in this case to the skewness of the $j$th component of $x$ (given that the components have mean zero and unit variance) it follows that an
off-diagonal element is zero only if at least one of the components has zero skewness. For example, consider the bivariate case $x = (x_1, x_2)^T$ with $x_1$ and $x_2$ being
independent and each having the discrete distribution with probability mass function $p(-0.5) = 0.8$ and $p(2.0) = 0.8$. This gives $\{\wcov_4(x)\}_{12} = 4.5$ and
$\{\wcov_4(x)\}_{jj} = 25.8125$ and hence a pseudo-correlation between $x_1$ and $x_2$ of 0.1743 even though they are independent.

\begin{figure}[tbph]
    \centering
    \includegraphics[width=0.8\textwidth]{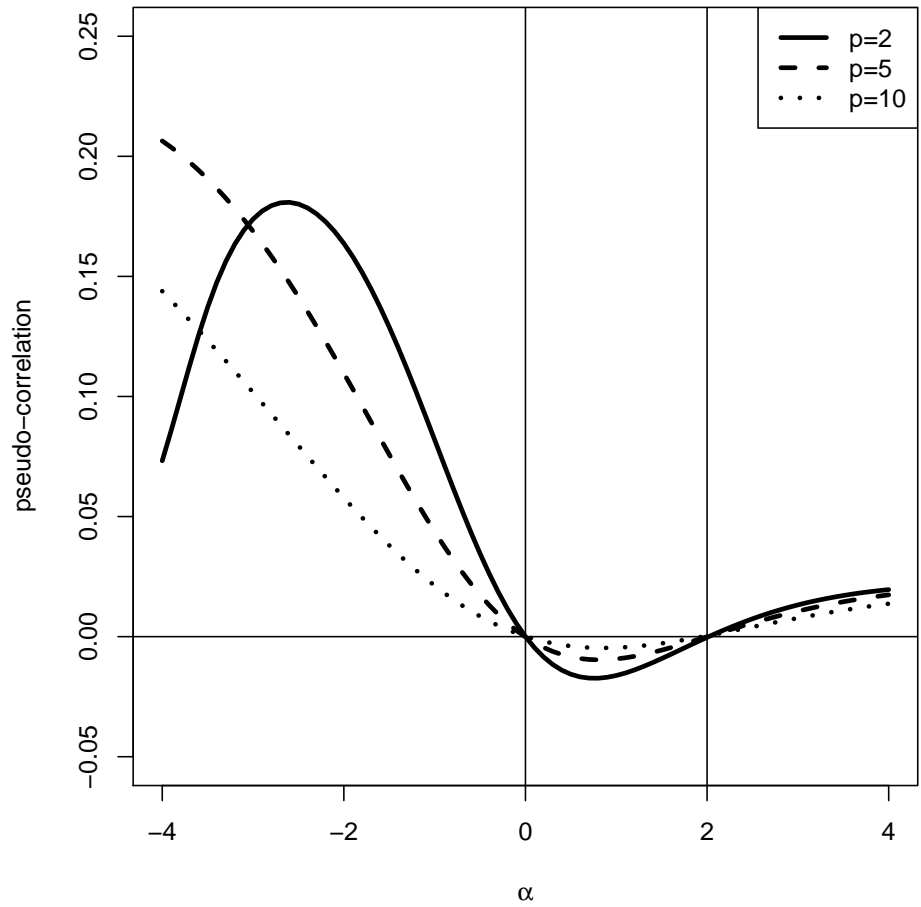}
    \caption{Value of the pseudo-correlation based on $\wcov_\alpha$. The vertical lines at 0 and 2 correspond to $\cov$ and $\wcov_2$ respectively.}
    \label{OffFig} 
\end{figure}

To demonstrate this idea further, Figure~\ref{OffFig} shows the pseudo-correlation obtained from $\wcov_\alpha(x)$ for
different values of $\alpha$ and $p$ in a setting where all $p$-components are mutually independent and each having a $\frac{1}{\sqrt{2}}(\chi_1^2-1)$ distribution.
Thus, the components have zero mean, unit variance and a skewness of $\sqrt{8} = 2.828$.  The results were obtained by taking the average, over 2000 repetitions,
of the sample version of $\wcov_\alpha(x)$ for samples of size 5000.

Figure~\ref{OffFig} clearly shows that the pseudo-correlations based $\wcov_\alpha(x)$ can be fairly large especially for
negative values of $\alpha$. Curiously, it is for $\alpha <0$ that $\cov_\alpha(x)$ has a more robust flavor since it corresponds to down-weighing values rather
than up-weighting values based on their original Mahalanobis distances. It can also be noticed that the pseudo-covariances are zero when $\alpha=0$, which
corresponds to the covariance matrix, and for $\alpha=2$.  The case $\alpha=2$, $\wcov_2(x)$ is sometimes referred to as a \emph{kurtosis matrix}, or as a matrix of fourth moments,
since it involves the fourth moments of $x$. It is known in general that $\wcov_2(x)$ is always diagonal whenever the components of $x$ are independent and possess
fourth moments, which is a key result needed to justify the well-known \emph{FOBI} algorithm in independent components analysis \citep{Cardoso:1989}.

The next counterexample utilizes the minimum volume ellipsoid (\emph{MVE}) estimators \citep{Rousseeuw:1986}. For a given $0 < h < 1$, the \emph{MVE} is defined as the ellipsoid
with the minimum volume covering at least $100h\% $ of the probability mass, say $(x-c)^T V^{-1}(x-c) \le 1$. The \emph{MVE} location functional is then taken to be the
center $c$ of this ellipsoid and the \emph{MVE} scatter functional $V_{MVE}(x; h)$ is taken to be proportion to $V$, with the constant of proportionality chosen so that
$V_{MVE}(x; h)$ corresponds to the covariance function when $x$ is multivariate normal.  For our admittedly artificial example, suppose the random vector $x=(x_1, x_2)^T$
has independent components with each component following a multinomial distribution with support 0, 1 and 2 and probabilities $0\cdot48$, $0\cdot45$ and $0\cdot07$ respectively.
For $h=0\cdot65$, the points covered by the \emph{MVE} can be shown to be $(0,0)^T$, $(1,0)^T$ and $(0,1)^T$, which then implies that
\[
V_{MVE}(x; 0\cdot65) = \frac{1}{\sqrt{3}}\left(
                     \begin{array}{cc}
                       4 & -2 \\
                       -2 & 4 \\
                     \end{array}
                   \right).
\]
Hence $V_{MVE}(x; 0\cdot65)$ yield as a robust pseudo-correlation of $-0\cdot5$ between the two independent components of $x$.

\subsection{Joint independence and symmetrization} \label{Section-Sym}
Of the scatter functionals considered so far, only $\cov$ and $\wcov_2$ are known to be diagonal whenever the components are
mutually independent.  \citet{OjaSirkiaEriksson:2006} refer to this property as the \emph{independence property} and discuss its
importance in independent components analysis.  Since we are to consider various notions of the independence property here, we refer
to this as the \emph{joint independence property}. That is,
\begin{definition} \label{indPropDef}
A scatter matrix $V(x)$ is said to have the joint independence property if, provided $V(x)$ exists,
\[
  V(x)= D(x),
\]
whenever $x$ has independent components and where D(x) is a positive diagonal matrix dependent on the distribution of $x$.
\end{definition}

A common feature of $\cov(x)$ and $\wcov_2(x)$ is that both can be expressed strictly in terms of
pairwise differences. Let $w$ and $v$ be two independent copies of $x$, then
\[
\cov(x)=\frac{1}{2}   E\left((w - v)(w - v)^T \right) \quad \mbox{and}
\]
\[
\wcov_2(x) =\frac{1}{2}  E\left((w -v)^T \cov(x)^{-1}(w -v) \cdot (w-v)(w-v)^T\right) -(p+2) \cov(x).
\]
In general, scatter functionals usually can not be expressed as a function of pairwise differences. On the other hand, given any scatter functional, one can generate its
\emph{symmetrized version} by simply applying the functional to pairwise differences.
\begin{definition} \label{SymVdef}
Let  $V(F_{x}) = V(x)$ be a scatter functional. Its symmetrized version is then defined to be
\[
  V_{sym}(x) := V(w - v),
\]
where $w$ and $v$ are independent copies of $x$.
\end{definition}
Symmetrized M-estimators are discussed in \citet{SirkiaTaskinenOja:2007}, while symmetrized S-estimators are discussed in \citet{RoelantVanAelstCroux:2009}.
The symmetrized version of the covariance matrix is simply $\cov_{sym}(x) = 2 ~\cov(x)$, whereas the symmetrized version of the kurtosis matrix is
$\wcov_{2,sym} = \wcov_2(x) + (p+2) \cov(x)$. As shown by Theorem 1 of \citet{OjaSirkiaEriksson:2006}, any symmetrized scatter matrix,
provided it exists, possesses the joint independence property. An open question, though, is whether these exist scatter matrices possessing the
joint independence property which cannot expressed as a function of pairwise differences.

\begin{figure}[tbph]
    \centering
    \includegraphics[width=0.8\textwidth]{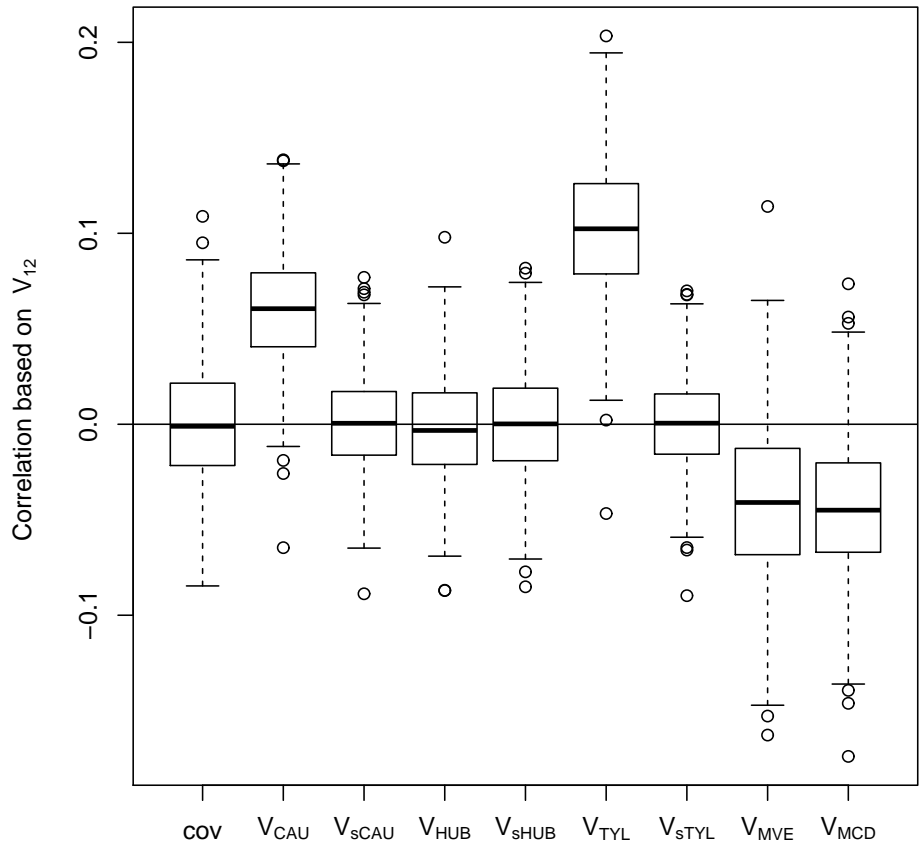}
    \caption{Box-plots of the pseudo-correlations for different scatter estimators arising from samples of size 1000, replicated 2000 times,
		from the $p=5$ dimensional random vector $x$ having mutually independent $\frac{1}{\sqrt{2}}(\chi_1^2-1)$ components.}
    \label{OffScatterFig} 
\end{figure}

Consider again the case where $x$ consists of independent $\frac{1}{\sqrt{2}}(\chi_1^2-1)$ components. For $p = 5$ and a sample size of 1000, Figure~\ref{OffScatterFig} shows the
box-plots of the simulated distribution, based upon 2000 repetitions, of the pseudo-correlations using on (i) the regular covariance matrix $\cov$, (ii) the  M-estimator derived as the maximum likelihood estimator of an elliptical Cauchy distribution $V_{CAU}$ \citep{KentTyler:1991}, (iii) the symmetrized version of $V_{CAU}$ denoted as $V_{sCAU}$,
(iv)  the M-estimator using Huber's weights $V_{HUB}$ \citep{Huber:1981},
(v) the symmetrized version of $V_{HUB}$ denoted $V_{sHUB}$, (vi) Tyler's shape matrix $V_{TYL}$ \citep{Tyler:1987},
(vii) the symmetrized version of $V_{TYL}$ denoted $V_{sTYL}$ (also known as D\"umbgen's shape matrix, \citet{Dumbgen:1998}),
(viii) the minimum volume estimator $V_{MVE}$ \citep{Rousseeuw:1986} and (ix) the minimum determinant estimator $V_{MCD}$ \citep{Rousseeuw:1986}. Throughout the paper, unless stated otherwise, the tuning constant for $V_{HUB}$ and $V_{sHUB}$ is taken to be 0.7 while for $V_{MVE}$ and  for $V_{MCD}$ is taken to be  $h=floor((n + p + 1)/2)$, where $n$ is the sample size and $p$ the dimension.
The box-plots are in agreement with our conjecture that in general only symmetrized scatter matrices have the joint independence property.

\subsection{Other independent structures} \label{Section:IndStr}

The joint independence property is weaker than property 3 of Lemma \ref{CovProp}.  That is, a scatter matrix $V(x)$ satisfying Definition \ref{indPropDef}
does not necessarily give $V_{jk}(x) = 0$ whenever $x_j$ and $x_k$ are independent.  For example, consider the kurtosis matrix $\wcov_2(x)$, which is
known to satisfy the joint independence property. Let $z_1, z_2$ and $z_3$ be mutually independent, each with zero mean and unit variance, and define
$x = (x_1, x_2, x_3)^T$, where $x_1 = z_1, x_2 = z_2$ and \mbox{$x_3 = 0.5(z_1+1)(z_2+1)z_3$}. It readily follows that $E(x) = 0$  and $\cov(x) = I_3$. Moreover, $x_1$ and $x_2$
are independent, but a simple calculation gives
\[ \{\wcov_2(x)\}_{12} = 0.25\{E(x_1^3)+2\}\{E(x_2^3) + 2\},\]
which is non-zero even for the case when $x$ has a symmetric distribution. Symmetrization does not help here since $\wcov_2$ is already symmetrized.  We conjecture that
no scatter matrix, other than the covariance matrix, satisfies property 3 of Lemma \ref{CovProp} in general.

As noted in \citet{TylerCritchletDuembgenOja:2009}, if more assumptions on the distribution of $x$ other than just independence are made, then unsymmetrized scatter
matrices can also yield zero pseudo-correlations. For example, if $x$ is symmetrically distributed about a center $\mu$, then any scatter functional $V(x)$, provided
it exist at $x$, is a diagonal matrix. This result immediately implies that a symmetrized scatter matrix has the joint independence property. In the following,
we state some further conditions under which independence implies a zero pseudo-correlation. The first result shows that symmetry can be slightly relaxed.


\begin{theorem} \label{diagVsym}
Let $x$ be a $p$-variate random vector with independent components. Furthermore, suppose $p-1$ components of $x$ are marginally
symmetric, i.e.\ for at least $p-1$ components, $x_j - \mu_j \sim -(x_j - \mu_j)$ for some $\mu_j$. Then any scatter matrix $V(x)$, provided
it exists at $x$, is a diagonal matrix.
\end{theorem}

Next, consider the case for which all $p$ components are $x$ are not necessarily mutually independent, but rather that the $p$-vector $x$ consists of independent blocks of
components. This means $x$ consists of $k \leq p$ sub-vectors $s_1, \ldots, s_k$ with dimensions $p_1,\ldots,p_k$, $\sum_{i=1}^k p_i=p$, such the $k$ sub-vectors
are mutually independent of each other. Such a setup arises for example in independent subspace analysis (ISA) \citep{NordhausenOja:2011}. We refer to this
property as the \emph{block independence property}.
\begin{definition} \label{indBlockPropDef}
Let $x$ have $k$ independent blocks with dimensions $p_1,\ldots,p_k$. The
 scatter matrix $V$ is said to have the block independence property if, provided $V(x)$ exists at $x$,
\[
  V(x)= B(x),
\]
where $B(x)$ is a block diagonal matrix with block dimensions $p_1,\ldots,p_k$.
\end{definition}
Clearly scatter matrices having the block independence property have the joint independence property. It is not clear though if the converse is true, i.e. whether the
joint independence property implies the block independence property. Nevertheless, as the corollary to the next theorem shows, symmetrization again assures that the
scatter matrix has zeros at the right places.
%
%
\begin{theorem}  \label{diagVIndBlock}
 Let $x=(x_1,\ldots,x_k)^T$ have $k$ independent blocks with dimensions $p_1,\ldots,p_k$. If at least $k-1$ blocks are symmetric in the sense that $x_i-\mu_i \sim -(x_i-\mu_i)$ where $\mu_i$ is the symmetry center of the $i$th block, then
any scatter matrix $V(x)$, provided it exists at $x$, will be block diagonal.
\end{theorem}

\begin{corollary} \label{diagVsymIndBlock}
Any symmetrized scatter matrix $V_{sym}(x)$ has the block independence \mbox{property.}
\end{corollary}

\section{Independent components analysis} \label{Section-ICA}
Independent components analysis (ICA) has become increasingly popular in signal processing and biomedical applications, where it is viewed as a practical
replacement for principal components analysis (PCA). ICA, in its most basic form, presumes that an observable random $p$-vector $x$ is a linear mixture of a
latent random $p$-vector $s$, with the components of $s$ being mutually independent.  Hence, the ICA model is commonly given as
\[
x=As,
\]
where $A$ is a full rank \emph{mixing} matrix. In order for the model to be identifiable, the \emph{signal} $s$ can have at most one normally distributed component. Even
then, the mixing matrix $A$ and signal $s$ are not completely identifiable, since $x$ can also be represented as $x = A_os_o$ where $s_o = PDs$ and $A_o = AD^{-1}P^T$,
with $P$ being a permutation matrix and $D$ being a full rank diagonal matrix.  This, though, is the only indeterminacy in the model. The primary goal in independent
components analysis (ICA) is to then find an \emph{unmixing} matrix $W$ such that $Wx$ has independent components.  Consequently, for some permutation matrix
$P$ and full rank diagonal matrix D, $W = A_o^{-1}$ and $Wx = s_o$. A general overview of ICA can be found, for example, in the often cited ICA book by
\citet{HyvarinenKarhunenOja:2001}.

Most approaches to ICA typically begin by first whitening the data using the sample covariance matrix. This is based on the observation that
\[
y = \cov(x)^{-1/2}  x =   Os,
\]
where $O$ is an orthogonal matrix whenever $s$ is viewed as a standardized signal, i.e.\ $\cov(s) = I_p$. After whitening the data, attention can then be focused on methods for
rotating the uncorrelated components of $y$ to obtain independent components. The approach of course presumes that $x$ possesses second moments. An obvious, though naive,
way to make this approach more robust would be to simply replace $\cov(x)$ with some robust scatter matrix $V(x)$. This is proposed, for example, by
\citep[][Section 14.3.2]{HyvarinenKarhunenOja:2001}, and by \citet{BalochKrimGenton:2005}, who recommend using the minimum covariance determinant (MCD) estimator.
However, in neither case is it noted that for such an approach to be valid either the signal $s$ must have a symmetric distribution, or more exactly to have
at most one skewed component, or the robust covariance must satisfy the independence property (\ref{indPropDef}), which e.g.\ is not satisfied by the MCD.  Problems
in practice, when simply replacing the regular covariance matrix with the MCD in the context of the popular fastICA method, have been noted
by \citet{BrysHubertRousseeuw:2005}. The reason such problems can arise is that if $V(x)$ does not satisfy  (\ref{indPropDef}), then $V(s)$ is not necessarily
diagonal and hence the signal may not correspond to any rotation of $y = V(x)^{-1/2} x$.

To quantitatively demonstrate the relevance of the independence property, we consider the bivariate case where $s$ has two skew independent components, the first
component having a $\chi_1^2$ distribution and the second component having a $\chi_2^2$ distribution, with both components being standardized to have mean zero
and unit variance. For this example, we use the ICA method proposed by \citet{OjaSirkiaEriksson:2006}. This ICA method requires two scatter (or shape)
matrices, say $V_1$ and $V_2$, with both satisfying the independence property. The method consists of using $V_1(x)$ to first whiten the data, giving $y = V_1(x)^{-1/2} x$,
and then performing a principal component analysis on $V_2(y)$. The resulting principal components of $y$ then correspond to the independent components. The
results are also the same when the roles of $V_1$ and $V_2$ are interchanged. For more details, see \citet{OjaSirkiaEriksson:2006}.

\begin{figure}[tbph]
    \centering
    \includegraphics[width=0.8\textwidth]{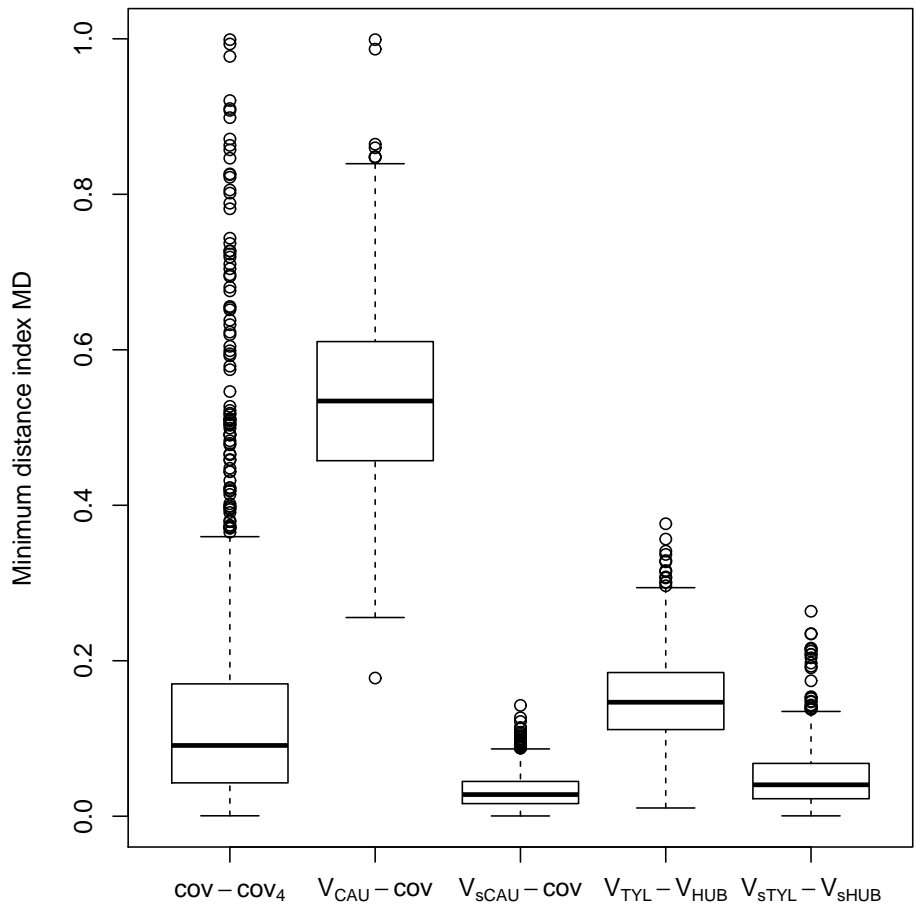}
    \caption{Box-plots of performance measure in $p = 2$ dimensions for the ICA method based on two scatter matrices, for various choices of the scatter matrices.
		The first component has a $\chi^2_1$ distribution and the second a $\chi^2_2$ distribution.}
    \label{ICAex} 
\end{figure}

A small simulation study was conducted using samples of size 1000 and with 1000 replications.
Since this ICA method is affine invariant, the choice of the mixing matrix $A$ has no effect on the performance of the method,
and so without loss of generality we take $A = I$.
Using the terminology established in the earlier sections,  we consider the following pairs of scatter matrices (i) $\cov$-$\cov4$ (ii) $V_{CAU}$-$\cov$,
(iii) $V_{sCAU}$-$\cov$, (iv) $V_{TYL}$-$V_{HUB}$, and (v) $V_{sTYL}$-$V_{sHUB}$. 
Case (iii) and (v) are the symmetrized version of (ii) and (iv) respectively.
Case (i) is already the same as its symmetrized version, and it corresponds to the classical FOBI method \citep{Cardoso:1989}.  Note that only for the
cases (i), (iii) and (v) do both scatter matrices satisfy the independence property. To measure the performance of the methods, we use the minimum distance
index MD, proposed in \citep{IlmonenNordhausenOjaOllila:2010}, which is defined to be
\[
MD(\hat W A) = \frac{1}{\sqrt{p-1}} \min_{P,D} ||PD \hat WA - I_p||,
\]
where $P$ is a permutation matrix and $D$ a diagonal matrix with non-zero entries. The range of the index is $[0,1]$, with 0 corresponding to an optimal recovery
of the independent components. Box-plots for the simulations are shown in Figure~\ref{ICAex}. The plots clearly show the relevance of the independence property here when there is more
than one asymmetric component, even in case (ii) which consists on only one scatter matrix without the independence property.
\\



\section{Observational regression through scatter matrices} \label{Section-ObsReg}

In this section we consider observation multivariate linear regression, that is linear regression for the case when the explanatory variables, as well as the responses, are
randomly observed rather than controlled. The classical multivariate linear regression model is then
\begin{equation}\label{RegModel}
y = \alpha + \mathcal{B}^T x + \epsilon,
\end{equation}
where $y$ is a $q$-dimensional response, $x$ is a $p$-vector of explanatory variables with distribution $F_x$, and $\epsilon \in \Re^q$ is a random error term, independent of $x$,
with distribution $F_\epsilon$. In this setting, interest usually is focused still on estimating the intercept vector $\alpha \in \Re^p$, the $p \times q$ slope matrix
$\mathcal{B}$ and perhaps the error variance-covariance matrix $\cov(\epsilon)=\Sigma_{\epsilon\epsilon}$ if it exists.

The standard least squares approach is well known to be highly non-robust, and so there have been numerous proposed
robust regression methods. One such method is based on the observation that if both $x$ and $\epsilon$ possess second moments, and if $\E(\epsilon) = 0$, then
\[
\mathcal{B} = \cov(x)^{-1} \cov(x,y), \quad \alpha = E(y) - \mathcal{B}^T\E(x), \quad \mbox{and} \quad \Sigma_{\epsilon\epsilon} = \var(y) - \mathcal{B}^T\cov(x)\mathcal{B},
\]
which corresponds to the population or functional version of the estimates arising from the least squares method. One can then generate a robust functional
version by again simply replacing the first two moments with robust versions of scatter and location.  That is, let $z=(x^T,y^T)^T$, which concatenates $x$ and $y$, and
consider the corresponding partitions of an affine equivariant location functional $\mu(z)$ and a scatter functional $V(z)$,
\[
\mu(z) = \left(
       \begin{array}{c}
         \mu_x  \\
         \mu_y \\
       \end{array}
     \right) \quad \mbox{and} \quad
V(z)=\left(
       \begin{array}{cc}
         V_{xx} & V_{xy} \\
         V_{yx} & V_{yy} \\
       \end{array}
     \right).
\]
If the distribution of $\epsilon$ is symmetric, then it has been observed in \citet{CrouxVanAelstDehon:2003} that the parameters $\alpha$ and $\mathcal{B}$ can also
be identified, even if no moments exist, through the equations
\[
\mathcal{B} = V_{xx}^{-1} V_{xy} \quad \mbox{and} \quad \alpha = \mu_y - \mu_x^T\beta,
\]
and so using the finite sample versions of $\mu(z)$ and $V(z)$ in the above relationship gives, under general regularity conditions, consistent
estimates of $\mathcal{B}$ and $\alpha$.

This approach was first proposed for univariate multiple regression by \citet{MaronnaMorgenthaler:1986} using $M$-estimators of multivariate location an scatter.
They note that this approach, unlike $M$-estimates of regression, yields bounded influence regression estimates.
This approach has also been studied for the Oja sign covariance matrix in \citet{OllilaOjaHettmansperger:2002}, for the Lift Rank Covariance Matrix
in \citet{OllilaOjaKoivunen:2003}, for S-estimators in \citet{CrouxVanAelstDehon:2003} and for the MCD in \citet{RousseeuwVanAelstVanDriessen:2004}.

The error variance $\Sigma_{\epsilon \epsilon}$ is not a robust functional itself, and is not identifiable when the error term does not have second moments.
Consequently, it is usually replaced by a robust scatter matrix for the residual term. Also, if $\epsilon$ does not have a symmetric distribution, then the
intercept term  $\alpha$ is confounded with the location of the error term \citep[Chapter 3 of][]{HettmanspergerMcKean:2011}. It has not been previously noted, though, how the relationship
$\mathcal{B} = V_{xx}^{-1} V_{xy}$ is affected by asymmetric error distributions. We first note that, due to the affine equivariance property of a scatter
(or shape) functional $V(z)$, this relationship always yields the proper equivariance properties for the slope parameters.
\begin{lemma} \label{RegEqui}
Let $y$ follow the regression model (\ref{RegModel}), assume that $V(z)$ exists with $V_{xx}$ being nonsingular, and denote $B(y,x)=V_{xx}^{-1} V_{xy}$.
Then $B(y,x)$ is regression, scale and design equivariant. That is, for $C_{p \times q}$, nonsingular  $M_{q \times q}$ and nonsingular $A_{p \times p}$,
\[ B(y+C^Tx,x) = B(y,x) + C,  \quad B(C^Ty,x) = B(y,x)C \quad \mbox{and} \quad B(y,Ax) = A^{-1}B(y,x).\]
\end{lemma}
Despite these equivariance properties, in order to obtain $B(x,y) = \mathcal{B}$, additional conditions on $V(z)$ are needed, which
as shown by corollary \ref{diagVsymIndBlock}, holds for symmetrized scatter/shape matrices.
\begin{theorem} \label{regSymV}
Let $y$ follow the regression model (\ref{RegModel}) and assume that $V(z)$ exists with $V_{xx}$ being nonsingular. Also, suppose
$V(z)$ satisfies the block independence property given by Definition \ref{indBlockPropDef}, then
$
B(y,x) = \mathcal{B}.
$
\end{theorem}


\begin{remark}
Consistency of the slope term under asymmetric errors has also been established for rank regression estimates and for $M$-estimates of regression. For details see for example \citet[Chapter 3 of][]{HettmanspergerMcKean:2011} and \citet[Chapter 4.9.2 of][]{MaronnaMartinYohai:2006} respectively.
\end{remark}

\begin{figure}[tbph]
    \centering
    \includegraphics[width=0.8\textwidth]{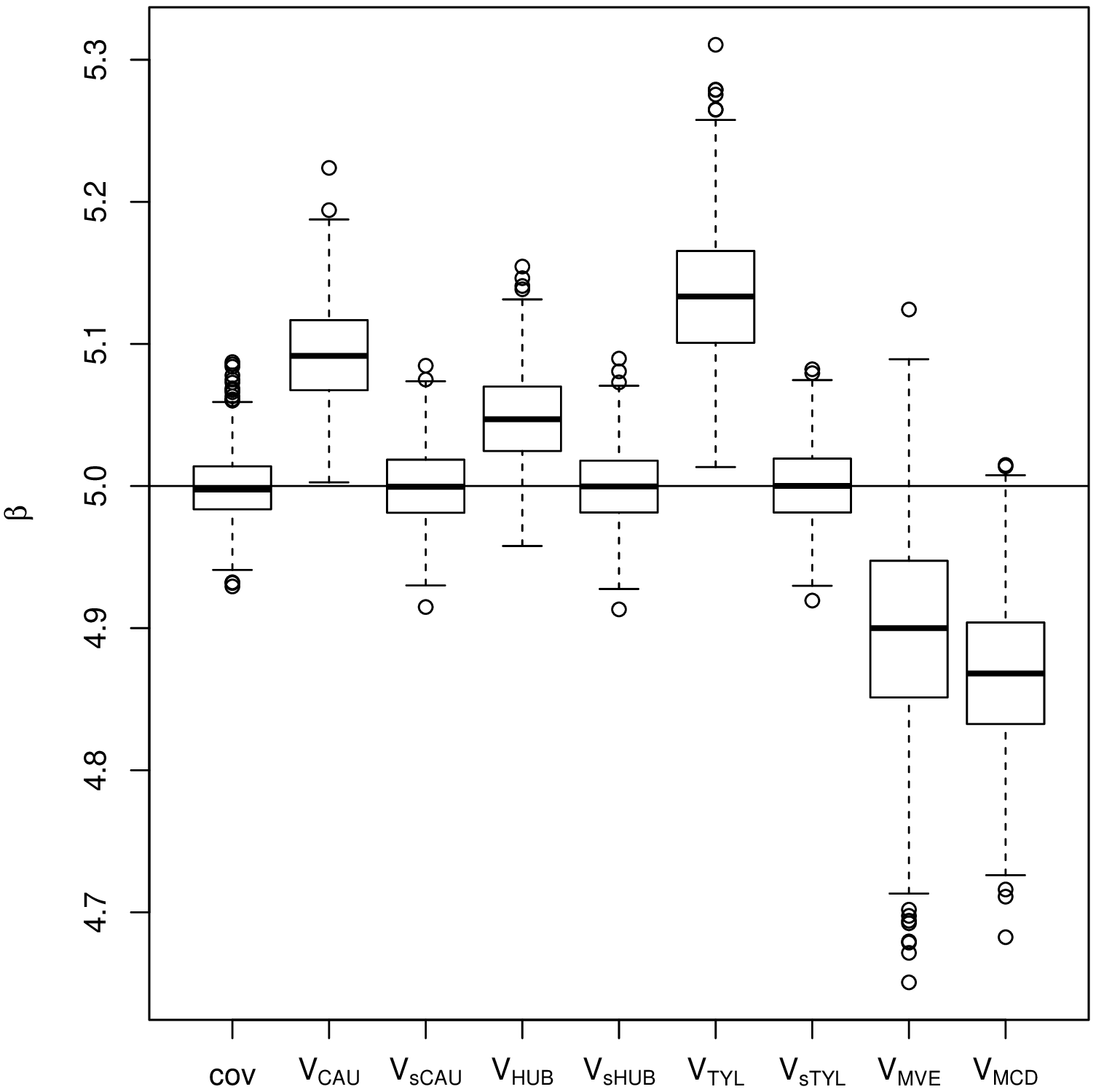}
    \caption{Comparing the performance the of symmetrized and not symmetrized scatter matrices for observational regression.}
    \label{REGex} 
\end{figure}

In order to demonstrate the necessity of symmetrization here whenever skewness is present in both $x$ and $\epsilon$, we conducted a simulation study for the model
\[
y= 5x + \epsilon,
\]
where $x$ has a log-normal distribution with shape parameter $\sigma=1$ standardized such that $E(x)=0$ and $\var(x)=1$ and $\epsilon$ has an exponential
distribution standardized to have $E(\epsilon)=0$ and $\var(\epsilon)=1$. For samples of size 2000, $\beta$ is estimated using (i) the regular covariance matrix $\cov$,
(ii)  M-estimator derived from as the maximum likelihood estimator of an elliptical Cauchy distribution $V_{CAU}$, (iii) the symmetrized version of $V_{sCAU}$ ,
(iv)  the M-estimator using Huber's weights $V_{HUB}$, (v) the symmetrized version of $V_{sHUB}$, (vi) Tyler's shape matrix $V_{TYL}$, (vii) the symmetrized
version of $V_{sTYL}$, (viii) the minimum volume estimator $V_{MVE}$ and (ix) the minimum determinant estimator $V_{MCD}$.
The results, based on 1000 replications and presented in Figure~\ref{REGex}, shows the severe bias when non-symmetrized scatter matrices are used.
which clearly shows that in this case the estimate for $\beta$ is severely biased when non-symmetrized scatter matrices are used.

\section{Graphical models} \label{Section-graphical}

The last method considered in this paper is graphical modeling for quantitative variables based on undirected graphs. In graphical models,
one is usually interested in those pairs of variables which are independent conditional on all the other variables, or, in graphical modeling terminology,
one is interested in those vertices (variables) which have no edges between them. In general, finding conditionally independent variables is challenging and
so finding variables with zero partial correlations often serves as a proxy.  In this section, we investigate the relationship between conditional independence
and robust versions of the partial correlation.

For \mbox{$p \geq 3$} random variables, consider the relationship between the variables $u$ and $v$ given $x$,
with $x$ containing the remaining $p-2$ variables. Denoting $y=(u,v)^T$, the partial variance-covariance matrix of $y$ given $x$ is given by
\[
\Sigma_{yy\cdot x}= \left(
                      \begin{array}{ccc}
                        \sigma_{11\cdot x} & ~ &\sigma_{12\cdot x} \\
                        \sigma_{21\cdot x} &  & \sigma_{22\cdot x} \\
                      \end{array}
                    \right),
										\]
where $\Sigma_{yy\cdot x}= \cov(y) - \cov(y,x) \cov(x)^{-1}\cov(x,y)$, which corresponds to the covariance matrix of the residuals
between the orthogonal projections of $u$ and $v$ onto the $p-2$-dimensional subspace spanned by $x$.
The corresponding partial correlation between $u$ and $v$ given $x$ is then simply
\[
\rho_{12\cdot x}=\frac{\sigma_{12\cdot x}}{\sqrt{\sigma_{11\cdot x} \sigma_{22\cdot x}}}.
\]
The partial correlation can also be expressed in terms of the precision or concentration matrix of the combined vector $z = (y^T,x^T)^T$. Specifically,
expressing the precision or concentration matrix of $z$ as $\Sigma_{z}^{-1}=\{\sigma^{ij}_z\}$, for $i,j=1,\ldots,p$,  where $\Sigma_{z}= \cov(z)$, one
obtains
\[
\rho_{12\cdot x}=- \frac{\sigma_z^{12}}{\sqrt{\sigma_z^{11} \sigma_z^{22}}},
\]
and hence $\rho_{12\cdot x} = 0$ if and only if $\sigma_z^{12} =0$.

For Gaussian graphical models, for which $z$ is presumed to be multivariate normal, conditional independence between $u$ and $v$ given $x$,
i.e.\ $u \perp v \mid x$, is equivalent to the partial correlation $\rho_{12\cdot x} = 0$. In general, conditional independence implies
a conditional correlation of zero, presuming the second moments exist, although the converse does not hold in general.  However, a perhaps lesser
known result is that conditional independence does not imply a zero partial correlation in general.  Some additional conditions are needed.  In particular,
if the regression of $y$ on $x$ is linear, then conditional independence implies a zero partial correlation, see Theorem 1 in \citet{BabaShibataSibuya:2004}.
Under such conditions, variables having zero partial correlations then serve as candidates for conditionally independent variables. When used in place of
conditional independence, zero partial correlations help provide a parsimonious understanding of the relationship between variables.

Robustness issues have been considered for graphical models, see for example \citet{Finegold:2011} and \citet{VogelFried:2011}. In both papers,
the emphasis is on finding pairs of variables for which a robust version of the partial correlations are zero. The approach used
in \citet{Finegold:2011} is a robust graphical lasso. The method uses a penalized maximum likelihood approach based on an elliptical $t$-distribution. The approach
advocated in \citet{VogelFried:2011} is a plug-in method based on using robust scatter matrices. They also study the asymptotic properties of the plug-in
method under elliptical distributions. Consequently, neither paper addresses conditional independence since conditional independence can never hold for variables
following a joint elliptical distribution other than the multivariate normal.

Outside the elliptical family, an important question worth addressing is under what conditions does conditional independence imply that the
the plug-in version of the partial correlation equals zero?  Since regression, i.e. the conditional mean of $y$ given $x$, is itself not a robust
concept and also is naturally related to covariances, the condition that regression be linear is not helpful here. We leave general conditions
under which conditional independence implies a zero robust partial correlation as an open question. We can, though, obtain results for
the following model
\begin{equation} \label{Graphmodel}
y = Ax + \epsilon,
\end{equation}
where $A$ is a non-random $2 \times (p-2)$ matrix, $\epsilon = (\epsilon_u, \epsilon_v)^T$, and $x$, $\epsilon_u$ and $\epsilon_v$ are mutually
independent. For this model, it readily follows that $u \perp v \mid x$. Also, if the first moments exist then the regression of $y$ on $x$ is linear.
Again, if one uses symmetrized scatter matrices than one obtains a plug-in version of the partial correlation which is equal to zero under this model.

\begin{theorem} \label{GraphModTheo}
Suppose model (\ref{Graphmodel}) holds, and assume that $V(z)$ exists and is nonsingular. Also, suppose
$V(z)$ satisfies the block independence property given by Definition \ref{indBlockPropDef}, then
$v^{12}_z=0,$ where $v^{jk}_z = \{V(z)^{-1}\}_{jk}$ is the $(j,k)$th element of the corresponding precision matrix.
\end{theorem}

\begin{figure}[tbph]
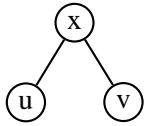

\centering
\pstree[levelsep=30pt]
{ \Tcircle{x} }
{
  \Tcircle{u} \Tcircle{v}
}
\caption{Graph used in the example.}
\label{GRAPH} 
\end{figure}
\begin{figure}[tbph]
    \centering
    \includegraphics[width=0.8\textwidth]{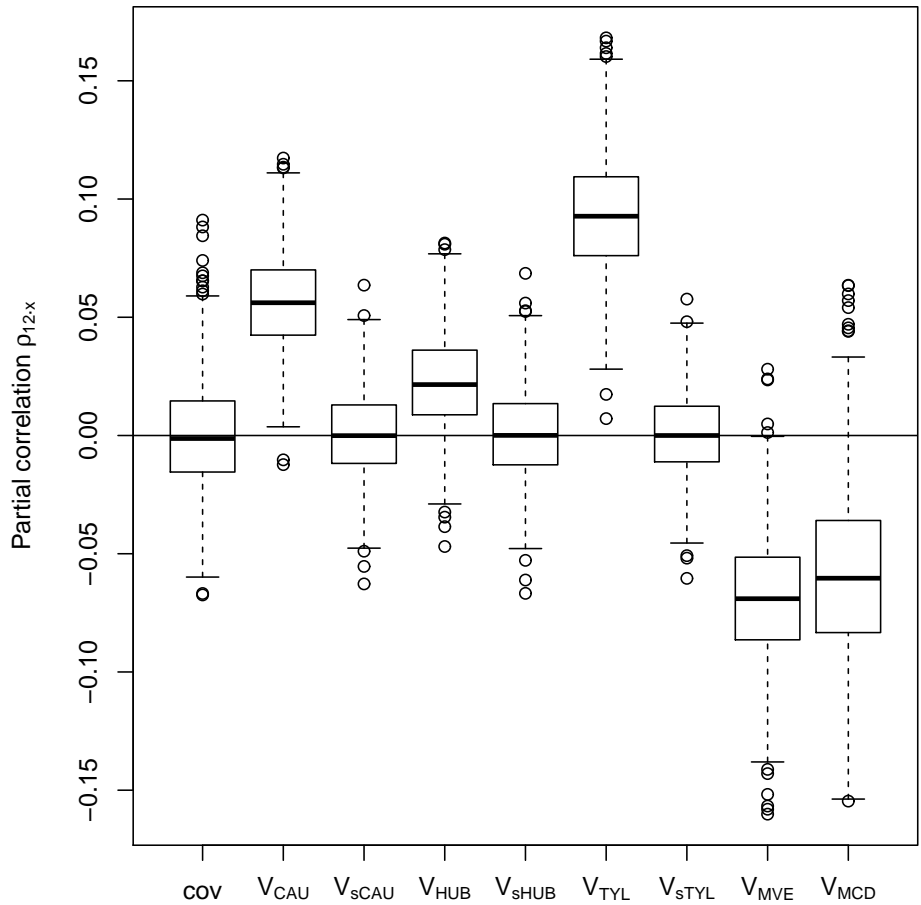}
    \caption{Comparing the performance the of symmetrized and not symmetrized scatter matrices for graphical modeling.}
    \label{GRAPHex} 
\end{figure}

As an example for illustrating Theorem~\ref{GraphModTheo}, consider the simple graphical model given in Figure~\ref{GRAPH},
where $u=4x+\epsilon_1$ and $v=5x+\epsilon_2$, with $x$ having a standard normal distribution, $\epsilon_1$ a log-normal distribution with shape parameter
$\sigma=1$ standardized such that $E(\epsilon_1)=0$ and $\var(\epsilon_1)=1$ and $\epsilon_2$ a $\chi^2_1$ distribution standardized to have
$E(\epsilon_2)=0$ and $\var(\epsilon_2)=1$. Using the same nine scatter matrices (i)-(ix) as in the previous section,  box plots for the plug-in partial
correlation of $u$ and $v$ given $x$ for sample of size 2000 based on 1000 replications are presented in Figure~\ref{GRAPHex}. Again, the
advantage to using symmetrized scatter/shape matrices is clearly shown.

\section{Computational aspects of symmetrization} \label{Section-Comp}

For various robust multivariate plug-in methods, we recommend symmetrized scatter matrices since they help protect against severe bias
whenever skew components are present. A drawback to using symmetrized scatter matrices, though, is that they are more computationally
intensive than their non-symmetrized counterparts. For a sample of size $n$, a symmetrized scatter matrix involves $n^2$ pairs. On the other hand, it does not
require an estimate of location since the difference is centered at the origin. Consequently, only those pairwise differences $x_i-x_j$ for which $i>j$ are
required for its computation and so the number of pairwise differences needed reduces somewhat to $n(n-1)/2$. Modern computers, though, have become so powerful
that computational cost should not deter the use of symmetrized scatter matrices when appropriate. Unfortunately, most robust scatter matrices implemented in
packages such as R do not allow the option of specifying the location vector, and so cannot be applied readily in computing symmetrized scatter matrices.
We hope the discussion in this paper will motivate future implementations of scatter matrices to include a fixed location option, as is the case in the
R packages ICS and ICSNP.

\begin{figure}[tbph]
    \centering
    \includegraphics[width=0.8\textwidth]{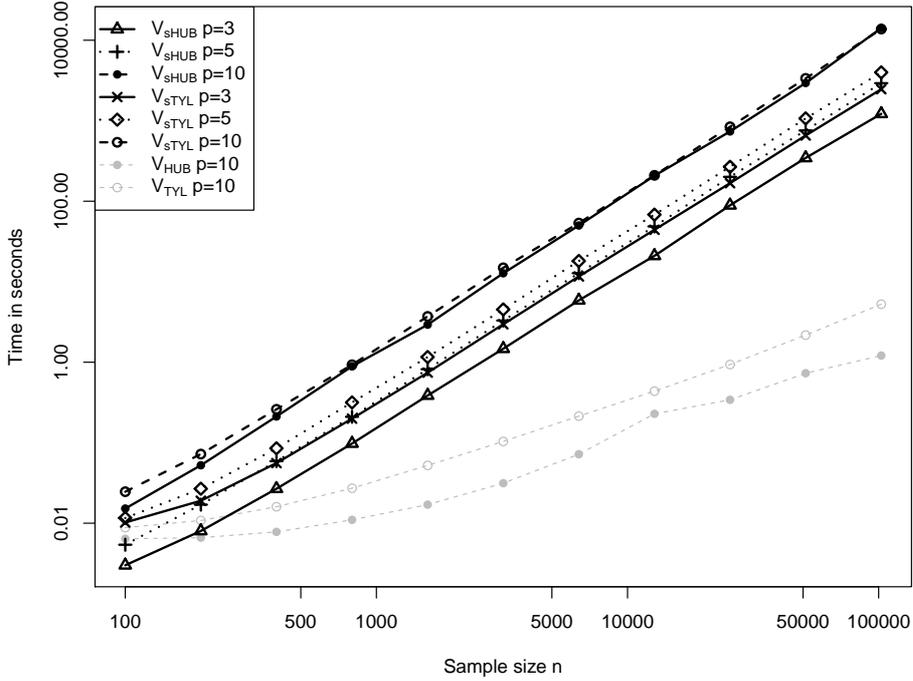}
    \caption{Average computation time in seconds for the symmetrized Tyler's shape matrix ($V_{sTYL}$) and for the symmetrized Huber M-estimator of scatter ($V_{sHUB}$)
		for various sample sizes $n$ and dimensions $p$. Both axes are given on a log-scale. The non-symmetrized version of the M-estimators are also given for $p=10$.}
    \label{CompT} 
\end{figure}

It may be difficult in general to develop algorithms which spread the computation of a scatter matrix over several cores. For $M$-estimates of scatter, though,
parallelization is possible. To see this, we note that when computing a symmetrized $M$-estimate of scatter $V_{sym}$ via the simple iteratively weighted least
squares algorithm, the update step is given by
\[
V_{sym,k+1} = \frac{2}{n(n-1)} \sum_{i=2}^{n} \sum_{j=1}^{i-1} w((x_i-x_j)^T V_{ym,k}^{-1}(x_i-x_j))(x_i-x_j)(x_i-x_j)^T,
\]
where $V_{sym,k}$ is the current value of the scatter matrix and $w(\cdot)$ is the weight function associated with the $M$-estimate. A simple way to compute the
symmetrized scatter matrix $V_{sym}$ which allows parallelization is to then set
\[
S_{k+1}^i = \sum_{j=1}^{i-1} w((x_i-x_j)^T V_{sym,k}^{-1}(x_i-x_j))(x_i-x_j)(x_i-x_j)^T,
\]
and so the iteration update for the symmetrized version becomes
{
\[
V_{sym,k+1}=  \frac{2}{n(n-1)} \sum_{i=2}^n S_{k+1}^i.
\]
}
To illustrate computation times, we considered the symmetrized version of Tyler's shape matrix $V_{sTYL}$, i.e.\ D\"umbgen's shape matrix,
implemented as \verb"duembgen.shape" in the R-package ICSNP and the symmetrized $M$-estimator of scatter using Huber's weights $V_{sHUB}$ implemented as \verb"symm.huber"
in the R-package SpatialNP. 
The average computing times
out of 5 runs for $N_p(0,\Sigma)$ data, where $\Sigma$ was randomly chosen, computed on a Intel(R) Xeon(R) CPU X5650 with 2.67GHz and 24GB of memory
running a 64-bit RedHat Linux are presented in Figure~\ref{CompT}. The figure shows that the computation time as of function of sample size is close to linear
when plotted on a log-log scale with a slope of approximately 2. Hence, the computation times are approximately of the order $n^2$. Also, for samples of
size $n =  500$ the computation times tend to be around one second, and that the symmetrized $M$-estimates are computationally feasible for even
fairly large sample sizes. As a comparison, for $p=10$, computation times for the non-symmetrized version of the M-estimators are also shown in the figure.

\section{Discussion} \label{Section-Discuss}

The goal of this paper has been to stress that some important or ``good'' properties of the covariance matrix do not necessarily carry over to affine
equivariant scatter matrices. Consequently, it is necessary to exercise some caution when implementing robust multivariate procedures based on the plug-in
method, i.e.\ when substituting a robust scatter matrix for the covariance matrix in classical multivariate procedures.  In particular, the validity of
some important multivariate methods require that the scatter matrix satisfy certain independence properties, which do not necessarily hold whenever
the components arise from a skewed distribution. Thus, we recommended the use of symmetrized scatter matrices in such situations, since
they are the only known scatter matrices which satisfy the independence property, Definition \ref{indPropDef}, or the block independence property, Definition
\ref{indBlockPropDef}.  We further conjecture that the only scatter matrices that satisfy these independence properties are those which can be expressed in
terms of the pairwise differences of the observation.

This paper has focused on the independence properties of scatter matrices. It would also be worth considering which scatter matrices, if any, possess the
additivity property of the covariance matrix, Lemma \ref{CovProp}.4. This property is relevant in factor analysis, in structural equation modeling, and in other
multivariate methods. For example, the factor analysis model is given by
\[
x = \Lambda f + \mu +\epsilon,
\]
where $f$ corresponds to $k<p$ latent factors and $\epsilon$ corresponds to a $p$-variate error term. $\epsilon$. The parameter $\mu$ represents a $p$-variate location
and $\Lambda$ corresponds to the $p \times k$ matrix of factor loadings (defined up to an orthogonal transformation). The standard factor analysis assumptions
are that the components of both $f$ are $\epsilon$ are mutually independent, and that $f$ and $\epsilon$ are also independent of each other. Furthermore, if the
first two moments exist, then is further assumed without loss of generality that $E(f)=0$, $\cov(f)=I_k$, $E(\epsilon)=0$ and $\cov(\epsilon)=D$, where $D$ is a
diagonal matrix with positive entries. Consequently, one can view such as factor analysis model as a reduced rank covariance model with an additive diagonal
term, i.e.\ as
\[
\cov(x) = \Lambda \Lambda^T + D.
\]
This decomposition is central to the classical statistical methods in factor analysis.  It is not clear though if one can define other scatter matrices so
that
\[
V(x) = \Lambda V(f)\Lambda^T + V(\epsilon),
\]
with both $V(f)$ and $V(\epsilon)$ being diagonal. Some robust plug-in methods for factor analysis and structural equation models have been
considered by \citet{PisonRousseeuwFilzmoserCroux:2003} and \citet{YuanBentler:1998a}.

%
\section*{Appendix: Proofs}
Let $J$ again represents a sign-change matrix, that is a diagonal matrix with diagonal elements of either $\pm 1$. Also, let $P$ represent a permutation matrix obtained
by permuting the rows and or columns of $I_p$.
\subsection*{Proof of Lemma~\ref{diagVess}}
For part 1, if $y \sim PJy$ for all $P$ and $J$ then $V(y)=V(Jy)=J V(y) J^T$ for all $J$, which implies all off-diagonal elements are zero.
Also, since $V(y)=V(Py)=PV(y)P^T$ for all $P$, it follows that all the diagonal elements are equal. Hence,
$V(y)= c_f I_p$, where $c_f$ is a constant depending on the density of $y$. Part 2 of the lemma then follows
from affine equivariance.


\subsection*{Proof of Theorem~\ref{diagVsym}}

Let $x=(x_1,\ldots,x_p)$ be a vector with independent components where $p-1$ components are marginally symmetric. Let $x_i$ be the component which
is not necessarily symmetric and let $J^i$ be any sign-change matrix for which the $i$th diagonal element is $+1$. Hence, $x \sim J^i x$ and due
to the affine equivariance of $V$ we have $V(x) = V(J^i x) = J^i V(x) J^i$ for any such $J^i$. This implies $V_{jk}(x) = - V_{jk}(x) = 0$ for $ j \ne k$
and hence $V(x)$ is a diagonal matrix.




\subsection*{Proof of Theorem~\ref{diagVIndBlock}}
Let $x=(x_1,\ldots,x_k)^T$ have $k$ independent blocks with dimensions $p_1,\ldots,p_k$, where all but the $i$th block are symmetric in the
sense that $-(x_j-\mu_j)\sim (x_j-\mu_j)$. Let $J_B$ denote a block sign-change matrix where the signs are changed according to blocks having
dimension $p_1,\ldots,p_k$ respectively. Also let $J_B^i$ denote a block sign-change matrix matrix where the $i$th diagonal block is $I_{p_i}$.
Since $x \sim J_B^i x$ for any  such $J_B^i$, it follows from the affine equivariance of $V$ that
$V(x) = V(J_B^i x) = J_B^i V(x) J_B^i$. This implies that off-diagonal block elements are zero and hence
$V(x)$ is block-diagonal with blocksizes $p_1,\ldots,p_k$.

\subsection*{Proof of Corollary~\ref{diagVsymIndBlock}}
Let $x$ have $k$ independent blocks and let $w$ and $v$ be independent identical copies of $x$. Then also $w-v$ has $k$ independent blocks.
Furthermore all blocks of $w-v$ are symmetric around the origin and so the corollary follows from Theorem~\ref{diagVIndBlock}.

\subsection*{Proof of Theorem~\ref{regSymV}}
Due to the equivariance properties stated in Lemma~\ref{RegEqui} it is sufficient to consider the case for which $\alpha=0$ and $\mathcal{B}=0$. For
this case $z=T(x^T,\epsilon)^T$ consists two independent blocks of dimensions $p$ and $q$, which by Theorem~\ref{diagVsymIndBlock}
implies $V(z)$ is block diagonal. Consequently, ${V}_{xy}=0$ and so $B(x,y)=0$.

\subsection*{Proof of Theorem~\ref{GraphModTheo}}
Let $z_o^T = (\epsilon^T, x^T)^T$. By Property \ref{indBlockPropDef}, it follows that
\[
V(z_o) = \left( \begin{array}{ccc}
\Delta & ~ & 0 \\
0 &  & M \\
 \end{array}
\right),
	\]
where $\Delta$ is a $2 \times 2$ diagonal matrix with positive diagonal terms, and $M$ is $(p-2) \times (p-2)$ positive definite symmetric matrix.
By affine equivariance, under model (\ref{Graphmodel}) it then follows that
\[
V(z) =
 \left( \begin{array}{ccc}
I & ~ & A \\
0 &  & I
 \end{array}
\right)
 \left( \begin{array}{ccc}
\Delta & ~ & 0 \\
0 &  & M
 \end{array}
\right)
 \left( \begin{array}{ccc}
I & ~ & 0 \\
A^T &  & I
 \end{array}
\right).
	\]
Taking the inverse gives

\begin{eqnarray*}
  V(z)^{-1} &=&  \left( \begin{array}{ccc}
I & ~ & 0 \\
-A^T &  & I
 \end{array}
\right)
 \left( \begin{array}{ccc}
\Delta^{-1} & ~ & 0 \\
0 &  & M^{-1}
 \end{array}
\right)
 \left( \begin{array}{ccc}
I & ~ & -A \\
0 &  & I
 \end{array}
\right) \\
   &=&  \left( \begin{array}{ccc}
\Delta^{-1} & ~ & -\Delta^{-1}A \\
A^T\Delta^{-1} &  & A^T\Delta^{-1}A + M^{-1}
 \end{array}
\right).
\end{eqnarray*}
Thus, $v_z^{12} = \{\Delta^{-1}\}_{12} = 0$.

\end{document}